\documentclass{article} 
\usepackage[margin=0.6in]{geometry}
\usepackage{graphicx,float}
\usepackage{cite}
\usepackage[justification=centering]{caption}
\usepackage{amsmath}
\usepackage{multirow}
\usepackage{tabularx}
\usepackage{subcaption}
\usepackage{authblk}

\title{\textbf{A Novel Gain Modeling Technique for LLC Resonant Converters 
based on The Hybrid Deep-Learning/GMDH Neural Network}}

\author[1]{Parham Mohammadi, Member, IEEE}
\affil[1]{Electrical Engineering and Computer Science Department, York University, Toronto, Ontario}

\date{} 

\begin{document}

\maketitle

\section{Introduction}
\begin{figure}[ht]
\centering
\includegraphics[width=4.5in]{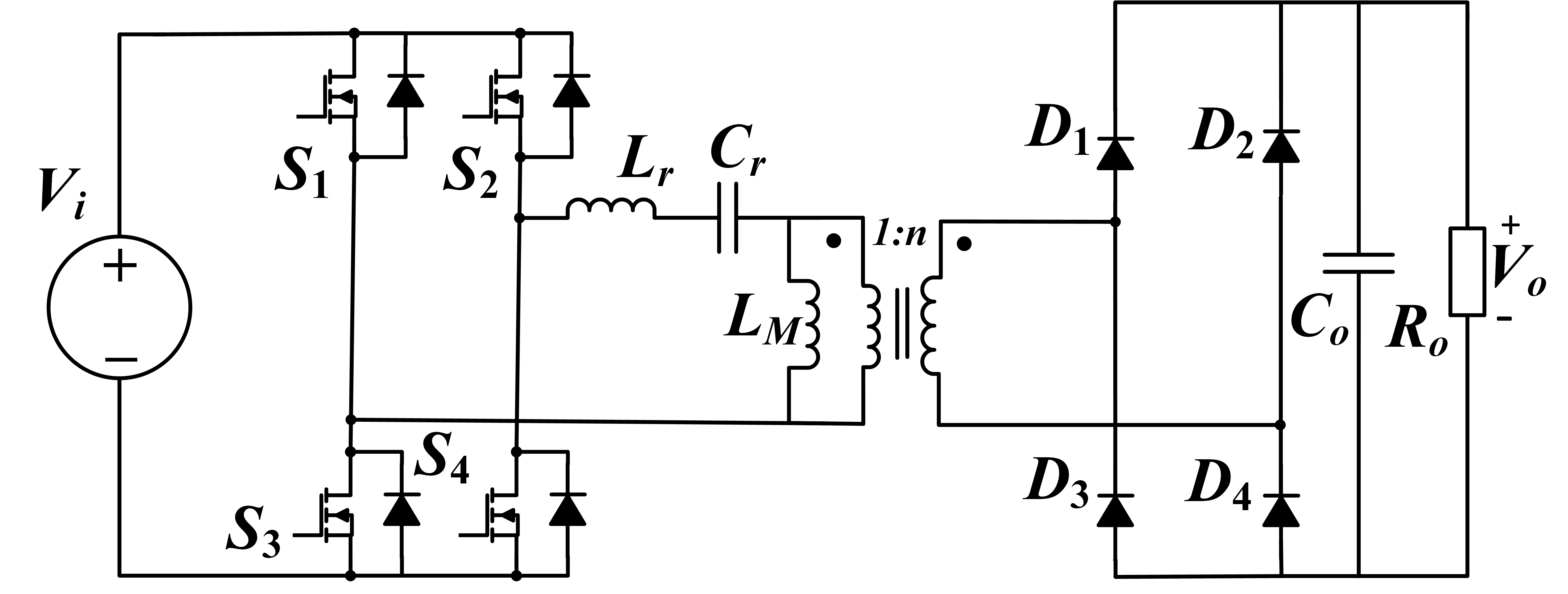}
\caption{Full-Bridge LLC Resonant Converter}
\label{fig:LLC}
\end{figure}

With the increasing global power demand in data centers and telecommunication, highly power-efficient energy management systems for data centers and telecom applications are crucial, both at the system and power conversion levels~\cite{YangChenQueensUniversity2017,Ahmed2020,Lu2020,Chen2018,Mu2016,Hayashi2015,Rizzolatti2019}.
The LLC resonant DC/DC converter, as shown in Fig.~\ref{fig:LLC}, is a widely used candidate in data center power conversion~\cite{YangChenQueensUniversity2017,Ahmed2020,Lu2020,Chen2018,Mu2016,Hayashi2015,Rizzolatti2019}. 

\begin{figure}[t]
\centering
\includegraphics[width=4in]{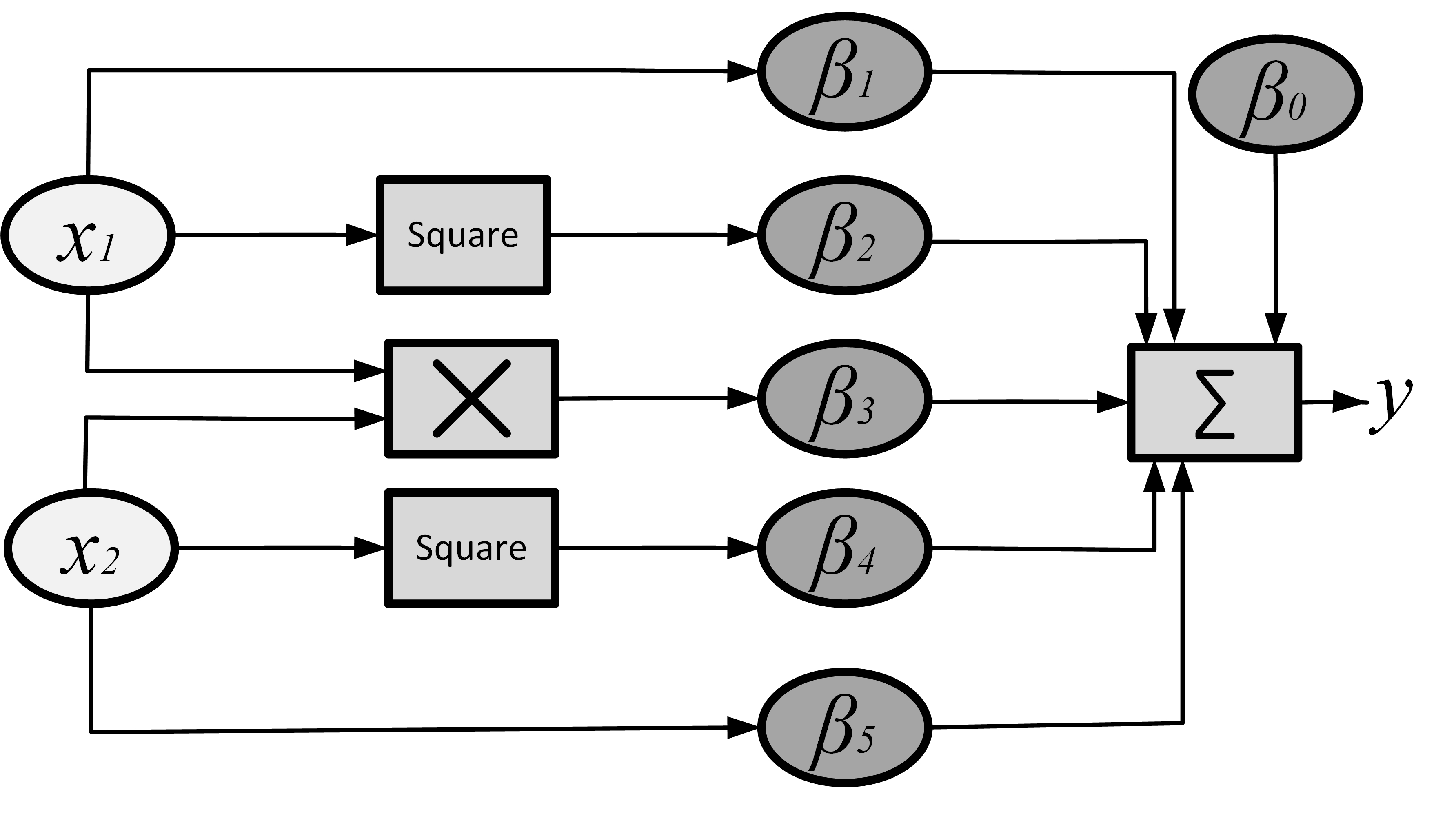}
\caption{Single neuron architecture in GMDH neural networks}
\label{fig:GMDH}
\end{figure}

\begin{figure*}[ht]
\centering
\includegraphics[width=6in]{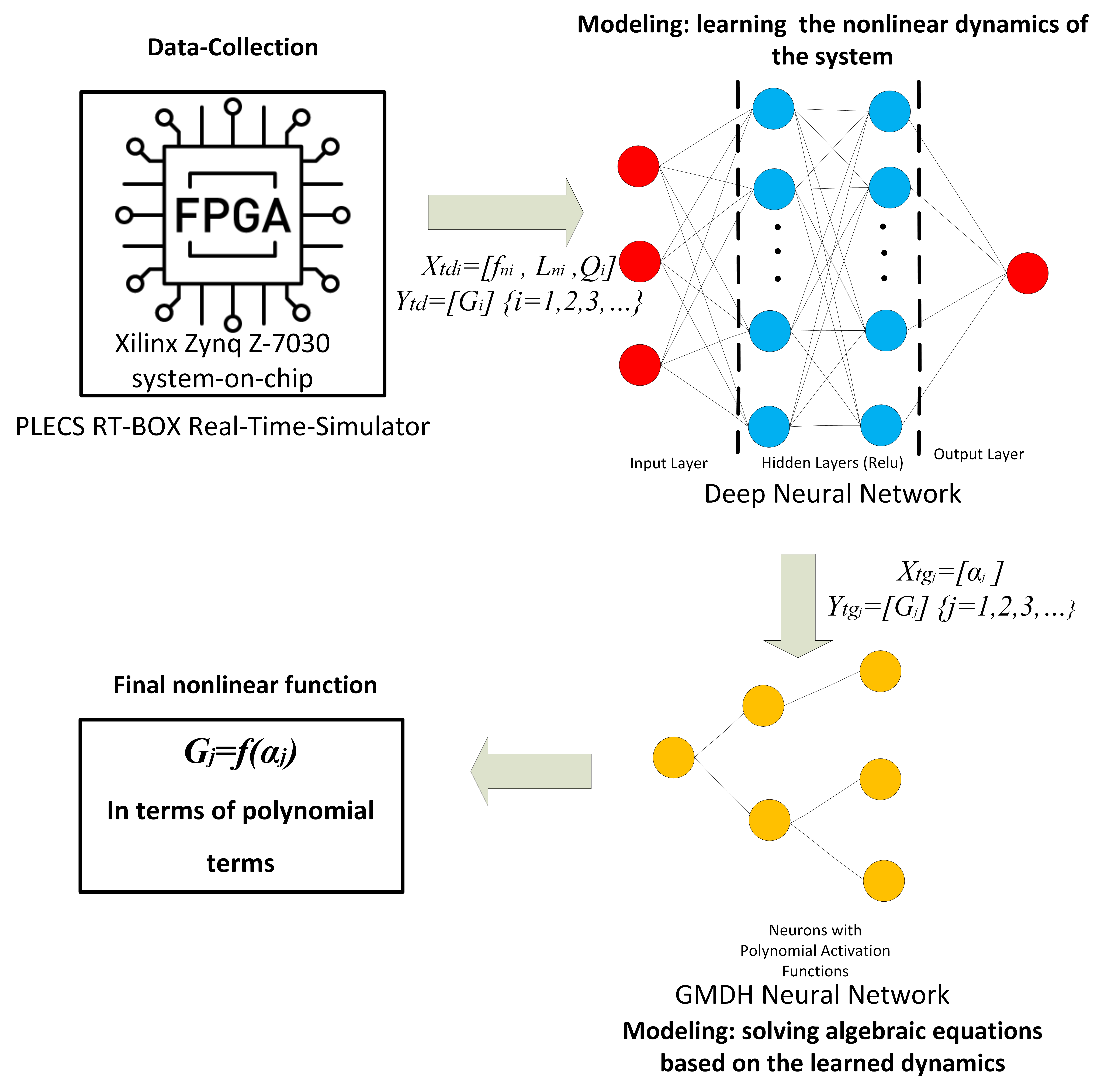}
\caption{Proposed hybrid deep-learning/GMDH neural network approach flowchart}
\label{fig:flowchart}
\end{figure*}

\par To optimize the LLC converter for achieving its highest efficiency, it is important that designers be able to model the converter accurately for a wide range of load conditions. Regarding the mathematical approach for determining the voltage gain of the LLC resonant converter, four major approaches have been discussed in the literature:
\vspace{0.25cm}
\begin{enumerate}
\item First harmonic approximation method (FHA method)~\cite{Steigerwald1988}.
\item Frequency domain analysis with partial time correction~\cite{Ivensky2011}.
\item Frequency domain with complete time correction~\cite{Liu2016}.
\item Time domain analysis (TDA)~\cite{Menke2020,Shen2020,Shafiei2017,Applications2020,Kumar2019,Glitz2019,Hong2010,Fang2011,Deng2014,Hu2015,Wei2019a}.
\end{enumerate}
\vspace{0.25cm}

While the FHA model is often used to calculate the gain of LLC resonant converters, it has two significant disadvantages. The first disadvantage is that it is assumed that the converter operates continuously in critical continuous mode, whereas in most instances, the LLC resonant converter does not. The second disadvantage is that the related AC resistance estimate is correct only when the converter is running in critical continuous mode, which is not always the case~\cite{Steigerwald1988,Ivensky2011,Menke2020,Shen2020,Shafiei2017,Applications2020,Kumar2019,Glitz2019,Hong2010,Fang2011,Deng2014,Hu2015}.

\par Frequency-domain with partial time and complete time-domain corrections have been presented to improve FHA's accuracy~\cite{Ivensky2011,Liu2016}. However, since their fundamental modeling structure is similar to that of the FHA method, they often do not calculate the voltage gain accurately~\cite{Menke2020,Shen2020,Shafiei2017,Applications2020,Kumar2019,Glitz2019,Hong2010,Fang2011,Deng2014,Hu2015,Wei2019,Wei2019b,Wei2019a}.

\par Research has been performed on modeling the LLC resonant converters in the time domain in order to overcome the limitations imposed by frequency-domain examination of the gain model~\cite{Menke2020,Shen2020,Shafiei2017,Applications2020,Kumar2019,Glitz2019,Hong2010,Fang2011,Deng2014,Hu2015,Wei2019a}. 
The model created using these approaches is much more accurate than those developed before. However, time-domain-based methods have high complexity, as they mostly involve solving highly nonlinear equations to obtain the resonant converter's gain model. Hence, constructing LLC resonant converters in this manner is very challenging~\cite{Wei2019,Wei2019b}.

\par Deep learning has been presented in the literature for use in power electronics~\cite{Lucia2020,Hajihosseini2020,Wang2019}. In particular, deep learning is capable of estimating nonlinear functions with very high accuracy. However, the disadvantage of deep learning is that, since the models developed by deep neural networks are spread over an extensive network, they cannot be used alone in a design process. 

\par The group method of data handling-type neural networks (Group Method of Data Handling-type Neural Networks) are a submodel of artificial neural networks. An approach that uses self-organizing networks provides positive results in a wide variety of domains~\cite{Heydari2021,Maric2010,Lin2015,Park2002,Iwasaki2004,Xiao2020,pukish2015polynet}. 
The architecture of an individual neuron of the GMDH neural networks is shown in Fig.~\ref{fig:GMDH}, where $\beta_{0}-\beta_{5}$ are the weights associated with each neuron's function that are generated via input data training. Each neuron's inputs are $x_{1},x_{2}$, and its output is $y$. The equations for each neuron may be derived as in (\ref{eq:G}).

\begin{equation}
y = \beta_{0} + \beta_{1}x_{1} + \beta_{2}x_{1}^2 + \beta_{3}x_{1}x_{2} + \beta_{4}x_{2}^2 + \beta_{5}x_{2}
\label{eq:G}
\end{equation}

In contrast to traditional neural networks used in deep learning, the model produced by GMDH neural networks is an algebraic equation based on a sequence of polynomial activation functions rather than a model distributed over a network. Equation (\ref{eq:G2}) is an infinite Volterra-Kolmogorov-Gabor (VKG) polynomial function~\cite{pukish2015polynet} which illustrates the final function form that a GMDH neural network generates, where $X=(x_{1},x_{2},...,x_{M})$ is the vector of input variables and $\Omega=(\omega_{0},\omega_{i},\omega_{ij},\omega_{ijk},...)$ is the vector of coefficients.

\begin{equation}
\begin{aligned}
Y =&\; \omega_{0} 
     + \sum_{i=1}^{M}\omega_{i}x_{i}
     + \sum_{i=1}^{M}\sum_{j=1}^{M}\omega_{ij}x_{i}x_{j} \\
   &\; + \sum_{i=1}^{M}\sum_{j=1}^{M}\sum_{k=1}^{M}\omega_{ijk}x_{i}x_{j}x_{k} + \dots
\end{aligned}
\label{eq:G2}
\end{equation}

As a result, the complexity of the model generated by the GMDH algorithm is much reduced when compared to traditional neural networks. However, in order to build an accurate but simple model using GMDH neural networks, these networks need a large quantity of input data.

\section{Proposed Hybrid Deep-Learning/GMDH Neural Network Approach for Modeling LLC Resonant Converter}

A new approach to accurately model the LLC resonant converter voltage gain based on GMDH neural networks and deep learning to predict resonant converters' gain is presented in this paper. 
It is well accepted that the inductor ratio $(L_{n})$, the quality factor $(Q)$, and the transformer turn ratio $(n)$ all influence the circuit characteristics of LLC resonant converters~\cite{Wei2019,Wei2019b}. 
In this paper these parameters are defined as in (\ref{eq:Params})--(\ref{eq:Params3}), where $L_{M}, L_{r}, C_{r}$, and $n$ are the transformer's magnetizing inductance, resonant inductor, resonant capacitor, and turn ratio, correspondingly.

\begin{equation}
L_{n}=\dfrac{L_{M}}{L_{r}}
\label{eq:Params}
\end{equation}

\begin{equation}
Q = \dfrac{ \sqrt{\dfrac{L_{r}}{C_{r}}} }{n^2 R_{o}}
\label{eq:Params2}
\end{equation}

\begin{equation}
f_{n} = \dfrac{f_{s}}{f_{r}} \quad \text{where} \quad f_{r} = \frac{1}{2\pi \sqrt{L_{r}C_{r}}}
\label{eq:Params3}
\end{equation}

Fig.~\ref{fig:flowchart} shows the flowchart that illustrates the process of developing the gain model for the LLC resonant converter.
$X_{td}=[(f_{n1}, L_{n1}, Q_{1}), \dots ]$ denotes the input vector, and $Y_{td}=[G_{1}, \dots]$ denotes the output vector utilized to train the deep neural network; these vectors were generated from simulations conducted by the real-time simulator. 
$X_{tg}=[\alpha_{1}, \dots]$ and $Y_{tg}=[G_{1}, \dots]$ are the input and output vectors produced by a deep neural network by mimicking the real-time simulator; these vectors are then employed to train the GMDH neural network. $\alpha$ is a feature designed to simplify the polynomials derived by the GMDH neural network, and it can be calculated as:

\begin{equation}
\alpha = \dfrac {1}{\sqrt{A^2 + B^2}},\quad
A = 1 + \dfrac{1}{L_{n}}\Bigl(1-\dfrac{1}{f_{n}^2}\Bigr),\quad
B = \bigl(f_{n}-\dfrac{1}{f_{n}}\bigr)\dfrac{1}{Q}.
\label{eq:Params4}
\end{equation}

The proposed idea combines the precision of deep learning and the simplicity of GMDH neural networks to produce this alternative model. 
Xilinx FPGA-based Real-Time Simulator is used to generate training and validation datasets for deep neural networks and GMDH neural networks. 
Deep learning is being utilized to create a massive quantity of data by mimicking the actual gain model of the LLC resonant converter generated by the real-time simulator. 
The deep learning results are then fed into GMDH neural networks to develop an accurate and straightforward model that can be used to calculate the gain of the resonant converter.

\section{Results and Performance}
\begin{figure*}[hb]
\centering
\includegraphics[width=4.5in]{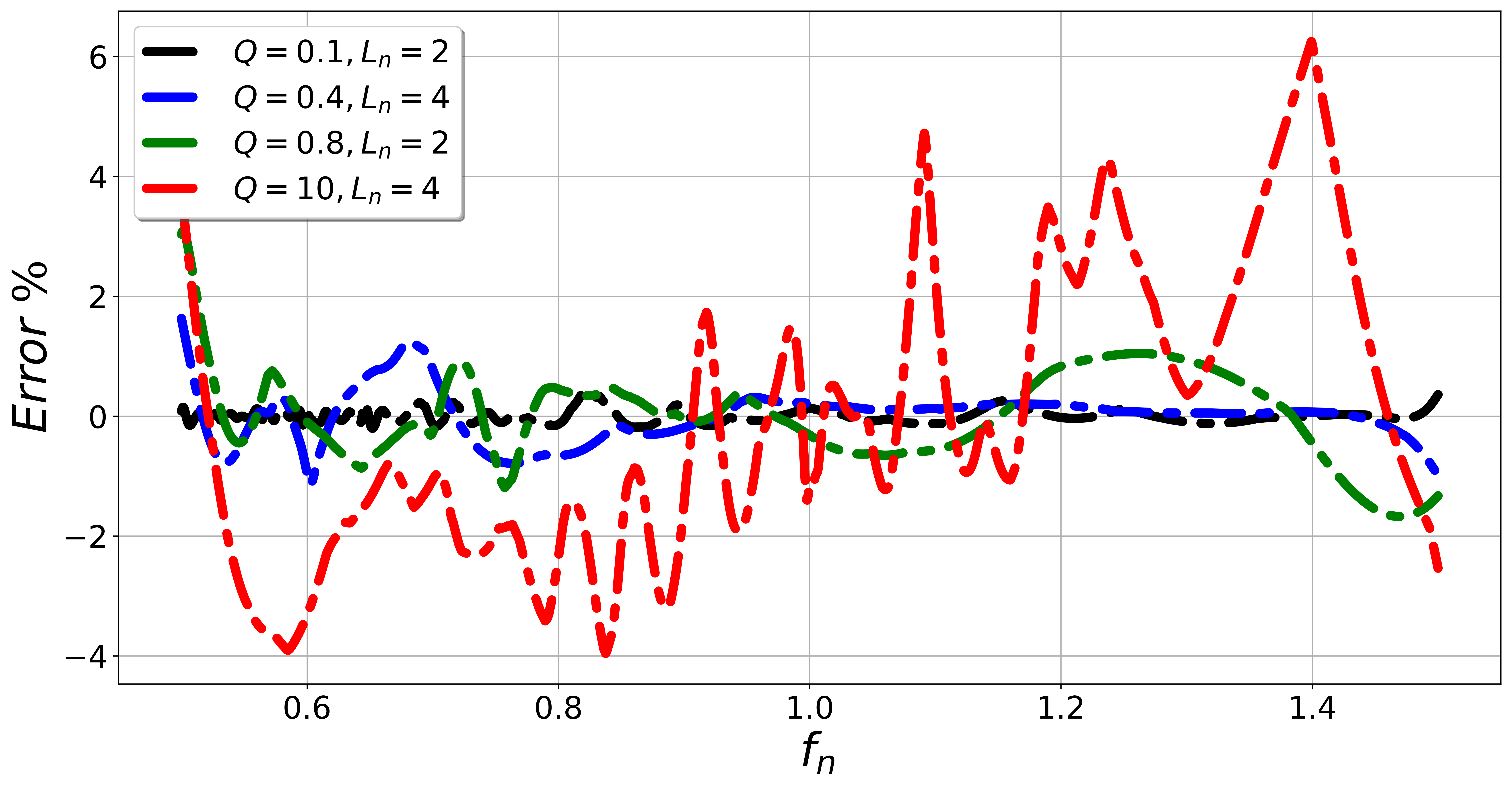}
\caption{Error between the results from the proposed hybrid model and real-time simulator results}
\label{fig:err}
\end{figure*}

To verify the proposed idea, a study of the proposed model has been performed on a wide range of $f_{n}$, from 0.5 to 1.5. Fig.~\ref{fig:err} depicts the error between the results generated by the proposed hybrid model and those obtained by the real-time simulator. The error is calculated by (\ref{eq:err}), where $G_{Hybrid}$ and $G_{RT}$ are the gains obtained by the proposed hybrid model and real-time simulator, respectively. Comparing the suggested model's error to other modeling techniques in the literature confirms its superiority, since other methods' error may reach $20\%$ in some instances~\cite{Wei2019,Wei2019b}.

\begin{equation}
\text{Error}_{i} = \dfrac{G_{Hybrid} - G_{RT}}{G_{RT}}
\label{eq:err}
\end{equation}

\begin{figure*}[ht]
\centering
\begin{subfigure}{0.35\textwidth}
  \centering
  \includegraphics[width=2.6in]{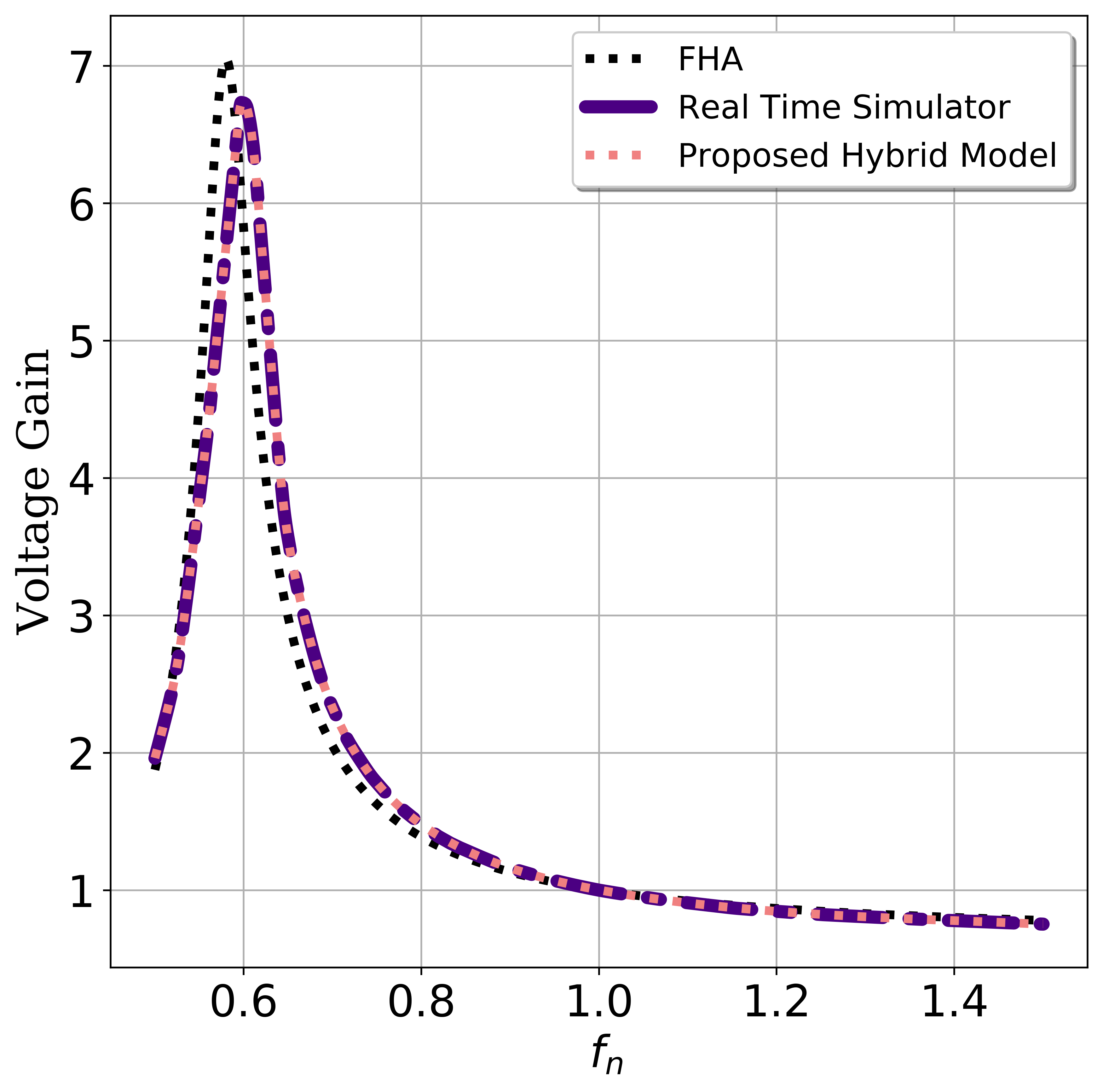}
  \caption{$L_{n}=2$ , $Q=0.1$}
  \label{fig:sub1}
\end{subfigure}
\begin{subfigure}{0.35\textwidth}
  \centering
  \includegraphics[width=2.6in]{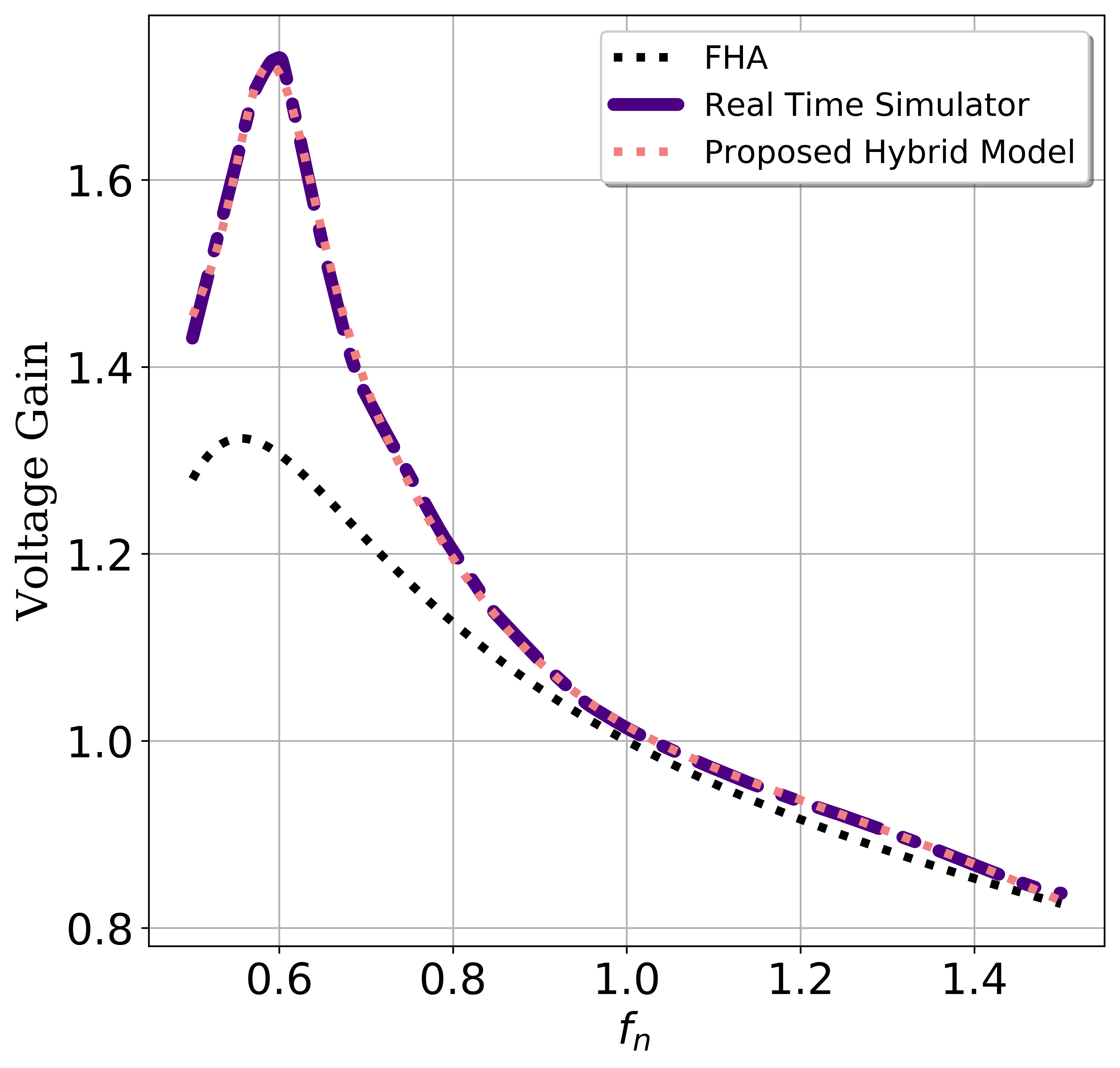}
  \caption{$L_{n}=4$ , $Q=0.4$}
  \label{fig:sub2}
\end{subfigure}
\begin{subfigure}{0.35\textwidth}
  \centering
  \includegraphics[width=2.6in]{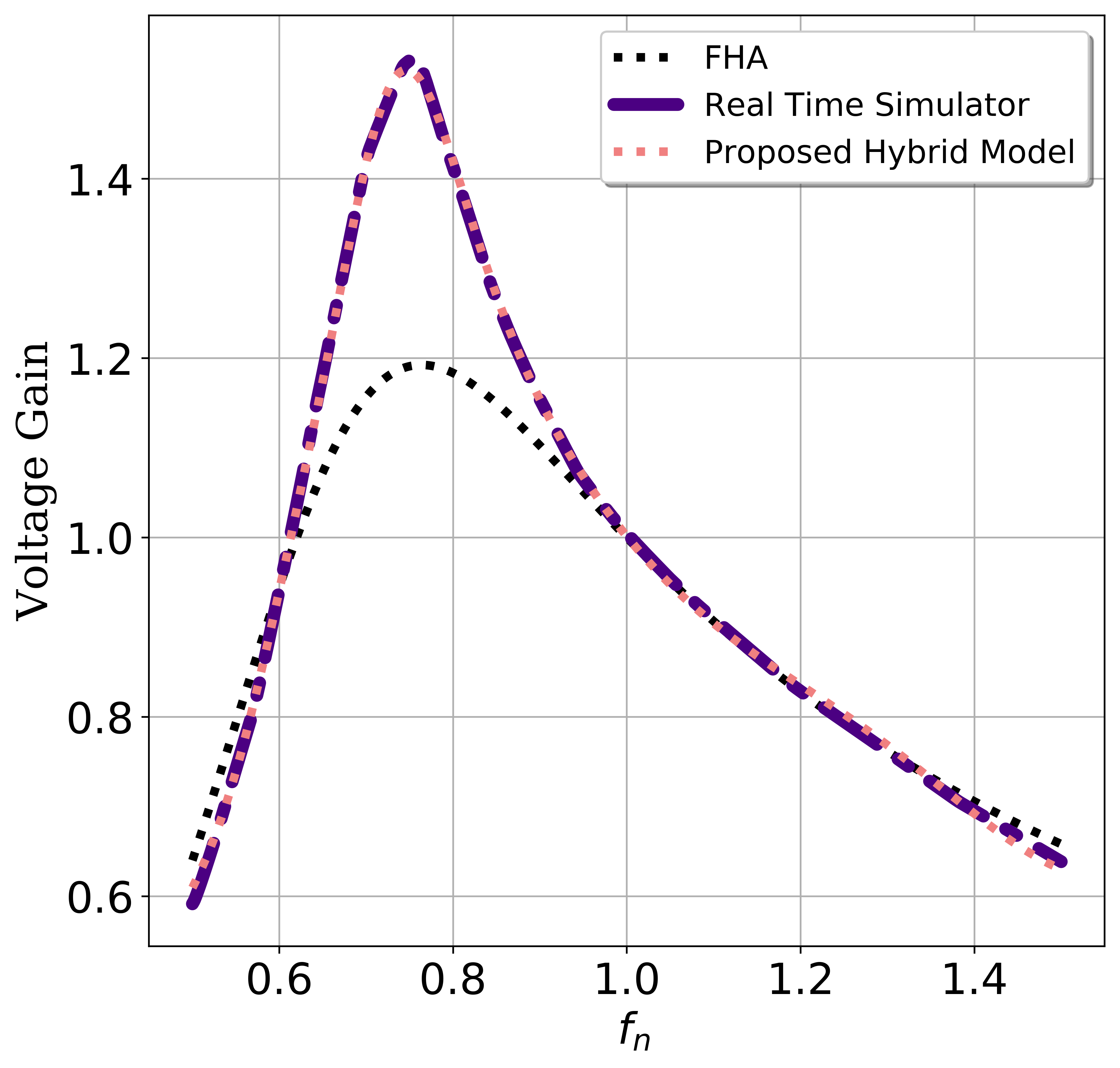}
  \caption{$L_{n}=2$ , $Q=0.8$}
  \label{fig:sub3}
\end{subfigure}
\caption{Comparison between FHA, Proposed Hybrid Method, and Real-Time Simulator}
\label{fig:Q}
\end{figure*}

Fig.~\ref{fig:Q} depicts the performance of the proposed hybrid method based on a validation dataset generated by the converter using the parameters given in Table~\ref{tab:train}.

\begin{table}[h]
\caption{Parameters of the Real-Time-Simulation Converter Used for Train and Validation Data Sets}
\label{tab:train}
\centering
\scalebox{1.1}{
\begin{tabularx}{230pt}{|c|c|c|}
\hline
\textbf{Parameters} & \textbf{Train Dataset} & \textbf{Validation Dataset} \\
\hline
Nominal $f_{s}$ & 20 kHz & 30 kHz \\
\hline
$L_{r}$ & 150 $\mu$H & 100 $\mu$H \\
\hline
$C_{r}$ & 0.4 $\mu$F & 0.267 $\mu$F \\
\hline
$n$ & 1 & 1 \\
\hline
$C_{o}$ & 220 $\mu$F & 220 $\mu$F \\
\hline
\end{tabularx}}
\end{table}

\par In summary, the proposed hybrid deep-learning/GMDH neural network approach for LLC resonant converter voltage gain modeling shows high accuracy and comparatively simpler implementation when contrasted with time-domain approaches. 

\vspace{1cm}


\end{document}